\documentclass[
 reprint,
 superscriptaddress,
 amsmath,
 amssymb,
 aps,
 pra,
]{revtex4-2}

\usepackage[dvipdfmx]{graphicx}
\usepackage{dcolumn}
\usepackage{bm}
\usepackage{color}
\usepackage{ulem}

\usepackage{ulem} 
\renewcommand{\sout}{\bgroup\markoverwith{\textcolor{red}{\rule[0.5ex]{2pt}{1pt}}}\ULon}

\begin{document}

\preprint{APS/123-QED}

\title{Optimal cavity design for minimizing errors in cavity-QED-based atom-photon\\ entangling gates with finite temporal duration}

\author{Takeru Utsugi}
\email{takeru.utsugi.qb@hitachi.com}
\affiliation{Department of Applied Physics, Waseda University, 3-4-1 Okubo, Shinjuku-ku, Tokyo 169-8555, Japan}
\author{Rui Asaoka}
\email{rui.asaoka@ntt.com}
\affiliation{Computer \& Data Science Laboratories, NTT Corporation, Musashino, Tokyo 180-8585, Japan}
\author{Yuuki Tokunaga}
\email{yuuki.tokunaga@ntt.com}
\affiliation{Computer \& Data Science Laboratories, NTT Corporation, Musashino, Tokyo 180-8585, Japan}
\author{Takao Aoki}
\affiliation{Department of Applied Physics, Waseda University, 3-4-1 Okubo, Shinjuku-ku, Tokyo 169-8555, Japan}

\date{\today}

\begin{abstract}
We investigate atom-photon entangling gates based on cavity quantum electrodynamics (QED) for a finite photon-pulse duration, where not only the photon loss but also the temporal mode-mismatch of the photon pulse becomes a severe source of error. 
We analytically derive relations between cavity parameters, including transmittance, length, and effective cross-sectional area of the cavity, that minimize both the photon loss probability and the error rate due to temporal mode-mismatch by taking it into account as state-dependent pulse delay.
We also investigate the effects of pulse distortion using numerical simulations for the case of short pulse duration.
We believe that these analyses are the first to suggest that a cavity has an optimal length for the atom-photon gate, providing a fundamental guideline for implementing quantum information processing.
\end{abstract}

\pacs{Valid PACS appear here}
\maketitle


Atom-photon gates are key components for universal quantum computation~\cite{nielsen2000quantum} based on cavity quantum electrodynamics (CQED)~\cite{reiserer2015cavity}, as these gates can be used to implement both atom-atom gates~\cite{xiao2004quantum,duan2005robust,PhysRevX.8.011018,daiss2021quantum} and photon-photon gates~\cite{duan2004scalable,hacker2016photon}, which are expected to enable the construction of scalable distributed quantum computation systems~\cite{reiserer2015cavity}. 
Recently, one of the entangling gates, a controlled phase flip (CPF) gate between atomic and photonic qubits in CQED systems, has been experimentally demonstrated~\cite{reiserer2014quantum,hacker2016photon,tiecke2014nanophotonic,PhysRevX.8.011018,daiss2021quantum}.

Previous theoretical investigations for atom-photon CPF gates have shown that the error probabilities are characterized by the internal cooperativity parameter $C_{\rm{in}}$ in the limit where the input-photon pulse duration is infinite~\cite{goto2005quantum,goto2010condition,asaoka2021requirements}. 
Since $C_{\rm{in}}$ is proportional to $\mathcal{F}/\mathcal{A}$ in typical cavities, where $\mathcal{F}$ is the finesse and $\mathcal{A}$ is the mode area~\cite{reiserer2015cavity,rosenblum2016extraction,goto2019figure}, in the long pulse limit, high-finesse cavities with small mode areas are required to suppress errors~\cite{vahala2003optical,kavokin2017microcavities,chang2018colloquium}.
On the other hand, we have previously shown that the error probabilities in the CPF gate cannot be characterized only by $C_{\rm{in}}$ when the photon pulse is of a finite duration~\cite{asaoka2021requirements}. This result suggests that the optimal cavity design should be reconsidered for finite pulse duration, which is required to achieve fast gate operation, and represents a more natural experimental description. 

In this Letter, we investigate the optimal cavity design for CQED-based CPF gates. 
In the case of finite pulse duration, the atom-photon gates have two severe error sources: the photon loss and the temporal mode-mismatch of the output photon pulse. 
Here, the dominant effect of temporal mode-mismatch can be considered as pulse delay.
In this approximation, we derive the relations between cavity parameters, including transmittance, length, and effective cross-sectional area of the cavity, that minimize both the photon loss probability and the error rate due to the pulse delay. 
We also perform numerical simulations of the dynamics of the CQED system and discuss the effect of pulse distortion when the pulse is short.
We note that the relationship between the gate error and the cavity length has yet to be discussed in previous studies~\cite{long1,long2,long3,long4,long5,long6,long7,long8,long9,long10}.
This is because CPF gates are typically analyzed in the long-pulse limit~\cite{note1}, where the gate error is independent of the cavity length.
Here, we investigate the finite-pulse case by extending the analytical treatment of the long-pulse limit.
We believe that our analyses with the temporal mode-mismatch error are the first to suggest that a cavity has an optimal length for the atom-photon gate, providing a fundamental guideline for implementing quantum information processing systems based on CQED.

\begin{figure}[tb]
\includegraphics[clip,width=80mm]{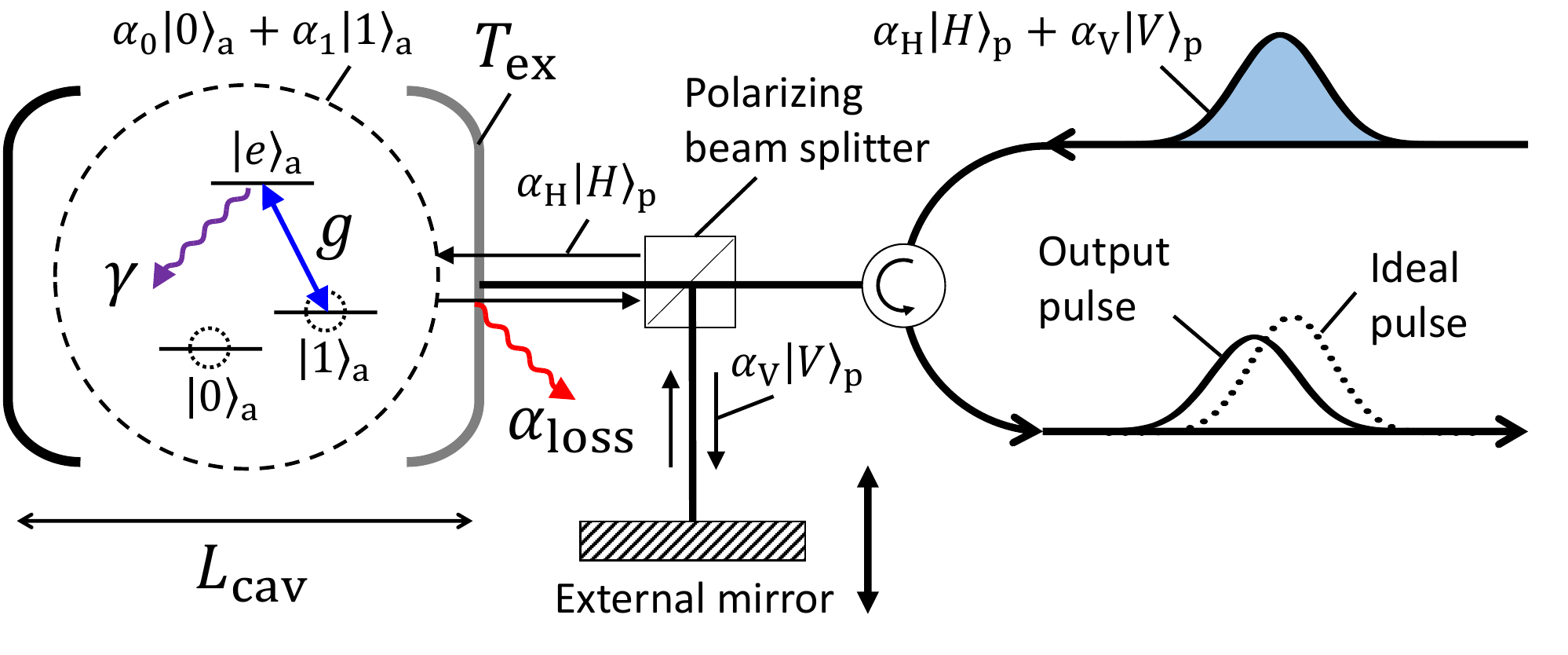}
\caption{Schematic of the atom-photon CPF gate. The photon carries qubit information represented by the polarization states $\alpha_{\rm{H}}|H\rangle_{\rm{p}}+\alpha_{\rm{V}}|V\rangle_{\rm{p}}$, and the two ground states of the trapped three-level atom in the cavity represent an atomic qubit $\alpha_{0}|0\rangle_{\rm{a}}+\alpha_{1}|1\rangle_{\rm{a}}$. 
The transition between $|1\rangle_{\rm{a}}$ and $|e\rangle_{\rm{a}}$ in the atom is resonant with the cavity.
Through the polarizing beamsplitter, the input photon having horizontal (vertical) polarization, $|H\rangle_{\rm{p}}$ ($|V\rangle_{\rm{p}}$), is directed to the one-sided cavity (the external mirror). The cavity length $L_{\rm{cav}}$, the cavity transmittance $T_{\rm{ex}}$, and the position of the external mirror are adjustable. The notation $g$ is the atom-cavity coupling constant, $\gamma$ is the atomic polarization decay rate, and $\alpha_{\rm{loss}}$ is the roundtrip loss rate of the cavity.}
\label{fig-model}
\end{figure}

Figure~\ref{fig-model} shows a setup to realize the atom-photon CPF gate~\cite{goto2005quantum,reiserer2014quantum,tiecke2014nanophotonic}.
An atom in a one-sided cavity has an excited state $|e\rangle_{\rm{a}}$, and ground states $|0\rangle_{\rm{a}}$ and $|1\rangle_{\rm{a}}$, which represent qubit bases. The transition between $|1\rangle_{\rm{a}}$ and $|e\rangle_{\rm{a}}$ is resonant with the cavity. A photon qubit is represented by its polarization $|H\rangle_{\rm{p}}$ and $|V\rangle_{\rm{p}}$. The initial state of the entire system is then written as
\begin{align}
	|\Psi_{\rm{in}}\rangle 
	= \left( \alpha_{0}|0\rangle_{\rm{a}} + \alpha_{1}|1\rangle_{\rm{a}} \right)\otimes 
	\left( \alpha_{\rm{H}}|H\rangle_{\rm{p}} + \alpha_{\rm{V}}|V\rangle_{\rm{p}} \right),
	\label{eq-Psi-in}
\end{align}
where $\alpha_{i}~(i\in\{0,1,H,V\})$ are complex amplitudes.

An input single-photon pulse is split into two paths using a polarizing beamsplitter so that only one polarization is incident on the cavity (we choose $|H\rangle_{\rm{p}}$ here). The other component is reflected by an external mirror.
We describe the output state after the atom-photon interaction via the cavity mode as
\begin{align}
	|\Psi_{\rm{out}}\rangle 
	&=L_{0}\alpha_{0}\alpha_{\rm{H}}|0H\rangle + \alpha_{0}\alpha_{\rm{V}}|0V\rangle \notag \\ 
	&+ L_{1}\alpha_{1}\alpha_{\rm{H}}|1H\rangle + \alpha_{1}\alpha_{\rm{V}}|1V\rangle,
	\label{eq-Psi-out}
\end{align}
where ${|jk\rangle=|j\rangle_{\rm{a}} \otimes |k\rangle_{\rm{p}}~(j\in\{0,1\},k\in\{H,V\})}$.
When ${L_{0}=-1}$ and ${L_{1}=1}$, the ideal CPF gate operation is realized.

Denoting the annihilation (creation) operators inside and outside of the cavity as $a_{\rm{c}} (a_{\rm{c}}^\dag)$ and $a_{\rm{p}} (a_{\rm{p}}^\dag)$, the Hamiltonian of the cavity path including the CQED system and a single-photon pulse is given by
\begin{align}
	\mathcal{H}
	=& \hbar \omega_{\rm{c}} a_{\rm{c}}^{\dagger}a_{\rm{c}} + \hbar \omega_{\rm{c}}|e\rangle_{\rm{aa}}\langle e| - \hbar\omega_{01}|0\rangle_{\rm{aa}}\langle 0|\notag \\
	+& \int^{\infty}_{-\infty}\!\!\!\!\!\!d\omega \hbar \omega a_{\rm{p}}^{\dag}(\omega)a_{\rm{p}}(\omega) 
	+ i\hbar g \left( a_{\rm{c}}^{\dagger} |1\rangle_{\rm{aa}}\langle e| - a_{\rm{c}} |e\rangle_{\rm{aa}}\langle 1|\right) \notag \\
	+& i \hbar \sqrt{\frac{\kappa_{\rm{ex}}}{\pi}}\int^{\infty}_{-\infty}\!\!\!\!\!\!d\omega[a_{\rm{c}}^{\dagger}a_{\rm{p}}(\omega)-a_{\rm{p}}^{\dagger}(\omega)a_{\rm{c}}],
	\label{eq-H}
\end{align}
where $\omega_{\rm{c}}$ and $\omega_{01}$ are the resonant frequencies of the cavity and the transition of $|0\rangle_{\rm{a}}$-$|1\rangle_{\rm{a}}$, respectively.
System parameters $g$ and $\kappa_{\rm{ex}}$ are the atom-cavity coupling constant and the external cavity field decay rate associated with the extraction of the cavity photon to the desired external mode. 
We also take into account the following dissipative processes: atomic spontaneous emission with a polarization decay rate $\gamma$ and unwanted cavity field decay due to the imperfection of the cavity with a rate $\kappa_{\rm{in}}$.
The total cavity decay rate is given by $\kappa=\kappa_{\rm{in}}+\kappa_{\rm{ex}}$.

The state of the atom and the photon having horizontal polarization is described as
\begin{align}
	|\psi_{\rm{H}}(t)\rangle
	&=|0\rangle_{\rm{a}}|0\rangle_{\rm{c}}\int_{-\infty}^{\infty} \!\!\!\!\!\!d\omega f_{0}(t,\omega) a_{\rm{p}}^{\dagger}(\omega)|0\rangle_{\rm{p}}  \notag \\
	&+\beta_{0}(t)|0\rangle_{\rm{a}}|1\rangle_{\rm{c}}|0\rangle_{\rm{p}} \notag \\
	&+|1\rangle_{\rm{a}}|0\rangle_{\rm{c}}\int_{-\infty}^{\infty} \!\!\!\!\!\!d\omega f_{1}(t,\omega) a_{\rm{p}}^{\dagger}(\omega)|0\rangle_{\rm{p}} \notag \\
	&+\beta_{1}(t)|1\rangle_{\rm{a}}|1\rangle_{\rm{c}}|0\rangle_{\rm{p}} \notag \\ 
	&+\beta_{e}(t)|e\rangle_{\rm{a}}|0\rangle_{\rm{c}}|0\rangle_{\rm{p}},
	\label{eq-psih}
	\end{align}
where $\omega$ denotes the frequency of the single-photon pulse, and the functions $f_{l}(t,\omega)~(l\in\{0,1\})$ and $\beta_{{m}}(t)~({m}\in\{0,1,e\})$ represent time-dependent probability amplitudes.

Here, we consider the case that the single-photon pulse is nearly resonant with the cavity, and we assume that there is no pure dephasing, which holds for typical CQED systems with neutral atoms~\cite{comm}.
{We calculate the dynamics of this system by the non-Hermitian Hamiltonian, which is enough to rigorously derive the gate fidelity equivalent to the master equation, in cases where the pure dephasing can be ignored~\cite{goto2005quantum}.}
Furthermore, we also consider the effect of the pure dephasing in Supplemental Material VII~\cite{sup}.
Thus, from the equations of motion of the state $|\psi_{\rm{H}}(t)\rangle$ and the input-output relations, we derive response functions in the form of amplitude reflection coefficients, as
\begin{align}
	&L_{0}(\Delta) = \frac{-\kappa_{\rm{ex}}+\kappa_{\rm{in}}+i\Delta}{\kappa_{\rm{ex}}+\kappa_{\rm{in}}+i\Delta},
	\label{eq-L0}\\
	&L_{1}(\Delta) = \frac{-\kappa_{\rm{ex}}+\kappa_{\rm{in}}+i\Delta+g^{2}/(\gamma+i\Delta)}{\kappa_{\rm{ex}}+\kappa_{\rm{in}}+i\Delta+g^{2}/(\gamma+i\Delta)},
	\label{eq-L1}
\end{align}
where $\Delta\equiv\omega_{\rm{c}} -\omega$ is the detuning from the resonant frequency (see Supplemental Material I~\cite{sup} for the derivation of the response functions).


In a typical one-dimensional cavity setup, $g$, $\kappa_{\rm{in}}$, and $\kappa_{\rm{ex}}$ are described as~\cite{reiserer2015cavity}
\begin{gather}
g =\sqrt{\frac{c\gamma}{2\tilde{A}_{\rm{eff}}L_{\rm{cav}}}},
\label{eq-g}\\
\kappa_{\mathrm{in}} =\frac{c\alpha_{\mathrm{loss}}}{4L_{\mathrm{cav}}},
\label{eq-kappa-in}\\
\kappa_{\mathrm{ex}} =\frac{cT_{\mathrm{ex}}}{4L_{\mathrm{cav}}},
\label{eq-kappa-ex}
\end{gather}
where $c$, $\tilde{A}_{\rm{eff}}$, $\alpha_{\rm{loss}}$, $T_{\rm{ex}}$, and $L_{\rm{cav}}$ are the speed of light in vacuum, the effective cavity mode area normalized by a scattering cross-section, the roundtrip loss rate, the coupling mirror transmittance, and the optical length of the cavity, respectively. 

We then rewrite Eqs.~(\ref{eq-L0}) and (\ref{eq-L1}) using Eqs.~(\ref{eq-g})-(\ref{eq-kappa-ex}) in the form
\begin{align}
	&L_{0}(\Delta) = \frac{-T_{\rm{ex}}+\alpha_{\rm{loss}}+i4L_{\rm{cav}}\Delta/c}{T_{\rm{ex}}+\alpha_{\rm{loss}}+i4L_{\rm{cav}}\Delta/c},
	\label{eq-L0re}\\
	&L_{1}(\Delta) = \frac{-T_{\rm{ex}}+\alpha_{\rm{loss}}+i4L_{\rm{cav}}\Delta/c+2/\left[\tilde{A}_{\rm{eff}}(1+{i\Delta/\gamma})\right]}{T_{\rm{ex}}+\alpha_{\rm{loss}}+i4L_{\rm{cav}}\Delta/c+2/\left[\tilde{A}_{\rm{eff}}(1+{i\Delta/\gamma})\right]}.
	\label{eq-L1re}
\end{align}

The conditions for an ideal phase flip in the long-pulse limit, i.e., the condition for $L_{0}=-1$ and $L_{1}=1$, are derived by substituting $\Delta=0$ into Eqs.~(\ref{eq-L0}) and (\ref{eq-L1}) or Eqs.~(\ref{eq-L0re}) and (\ref{eq-L1re}) as~\cite{asaoka2021requirements}:
\begin{equation}
2C_{\rm{in}}\gg\frac{\kappa_{\rm{ex}}}{\kappa_{\rm{in}}}\gg1,
\label{eq-basic1}
\end{equation}
or
\begin{equation}
\frac{2}{\tilde{A}_{\rm{eff}}}\gg T_{\rm{ex}}\gg\alpha_{\rm{loss}},
\label{eq-basic2}
\end{equation}
where $C_{{\rm{in}}}\equiv g^{2}/(2\gamma\kappa_{\rm{in}})=1/(\tilde{A}_{\rm{eff}}\alpha_{\rm{loss}}$) is an internal cooperativity~\cite{rosenblum2016extraction,goto2019figure}.

\begin{figure}[tb]
\includegraphics[clip,width=80mm]{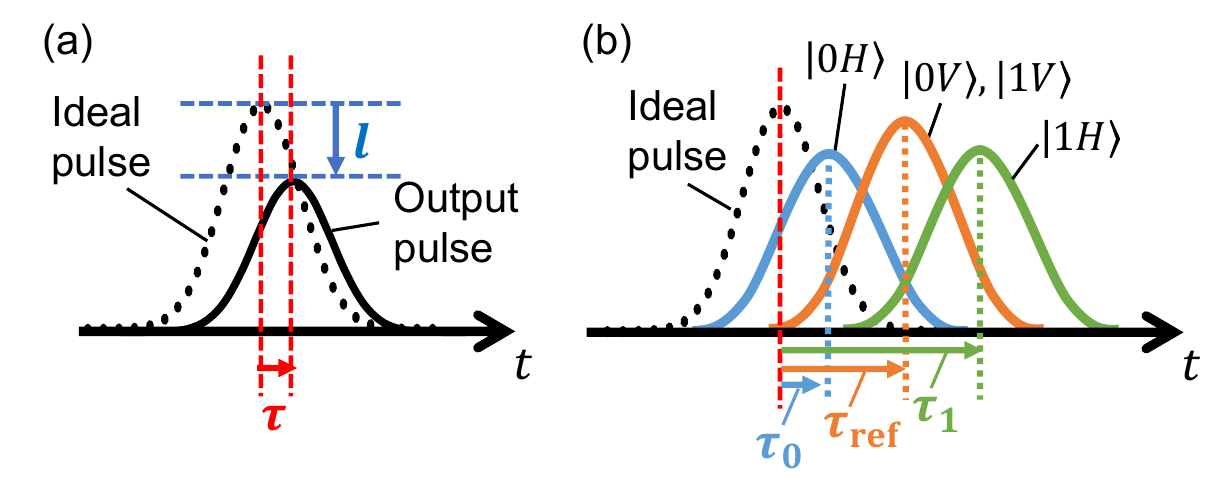}
\caption{Schematic of error sources: (a) the loss $l$ and delay $\tau$ of the output photon pulse and (b) the output pulse delay depending on the atomic and photonic states and the external mirror position.}
\label{fig-LossDelay}
\end{figure}

To analytically estimate the error probability, we expand the response functions by $\Delta$ as
\begin{align}
	L_{0}(\Delta) &= \frac{\alpha_{\rm{loss}}-T_{\rm{ex}}}{\alpha_{\rm{loss}}+T_{\rm{ex}}} + i\frac{8L_{\rm{cav}}T_{\rm{ex}}}{c(\alpha_{\rm{loss}}+T_{\rm{ex}})^{2}}\Delta + O(\Delta^{2}),
	\label{eq-L0approx2}\\
	L_{1}(\Delta) &= \frac{\alpha_{\rm{loss}}-T_{\rm{ex}}+2/\tilde{A}_{\rm{eff}}}{\alpha_{\rm{loss}}+T_{\rm{ex}}+2/\tilde{A}_{\rm{eff}}}\notag\\
	 &+ i \frac{4T_{\rm{ex}}(2\tilde{A}_{\rm{eff}}L_{\rm{cav}}\gamma-c)}{c\gamma(\alpha_{\rm{loss}}+T_{\rm{ex}}+2/\tilde{A}_{\rm{eff}})^{2}}\Delta +O(\Delta^{2}).
	\label{eq-L1approx2}
\end{align}
In both equations, the first terms are real and represent the photon loss, i.e., they contribute to the reduction in the norm of the state function.
The second terms are imaginary and describe the pulse delay.
A schematic diagram of the reduction in the norm and the pulse delay is shown in Fig.~\ref{fig-LossDelay}(a).

When an incident pulse is sufficiently long that the terms higher than the second order are negligible (the validity of this approximation and the effect of the higher order terms are discussed later), the photon loss probabilities when the atom is in $|0\rangle_{\rm{a}}$ or $|1\rangle_{\rm{a}}$ are derived as~\cite{goto2010condition}:
\begin{align}
	&l_{0} \equiv 1-L_{0}(0)^{2}=1-\left(\frac{\alpha_{\rm{loss}}-T_{\rm{ex}}}{\alpha_{\rm{loss}}+T_{\rm{ex}}}\right)^{2},
	\label{eq-l0}\\
	&l_{1} \equiv 1-L_{1}(0)^{2}=1-\left(\frac{\alpha_{\rm{loss}}-T_{\rm{ex}}+2/\tilde{A}_{\rm{eff}}}{\alpha_{\rm{loss}}+T_{\rm{ex}}+2/\tilde{A}_{\rm{eff}}}\right)^{2}.
	\label{eq-l1}
\end{align}
These are equivalent to the norm reduction of the states $|0H\rangle$ and $|1H\rangle$, which exist even in the long-pulse limit.

The error due to the photon loss is classified twofold:
the reduction in the norm for the total state function, and the unwanted rotation of the state vector due to imbalances in the norm reduction for the bases.
The former and latter error probabilities are minimized on the condition of $l_{0}+l_{1}$ and $|l_{0}-l_{1}|$ being minimized, respectively.


As shown in Supplemental Material II~\cite{sup}, the pulse delays for the inputs of the $|0H\rangle$ and $|1H\rangle$ states are given by
\begin{align}
	\tau_{0} &\equiv \frac{i}{L_{0}(0)}\left.\frac{dL_{0}(\Delta)}{d\Delta}\right|_{\Delta=0}=\frac{8L_{\rm{cav}}T_{\rm{ex}}}{c(T_{\rm{ex}}^{2}-\alpha_{\rm{loss}}^{2})},
	\label{eq-tau0}\\
	\tau_{1} &\equiv \frac{i}{L_{1}(0)}\left.\frac{dL_{1}(\Delta)}{d\Delta}\right|_{\Delta=0}\notag\\
	&=-\frac{T_{\rm{ex}}(2\tilde{A}_{\rm{eff}}L_{\rm{cav}}\gamma-c)}{c\gamma[\alpha_{\rm{loss}}+1/\tilde{A}_{\rm{eff}}-\tilde{A}_{\rm{eff}}(T_{\rm{ex}}^{2}-\alpha_{\rm{loss}}^{2})/4]}.
	\label{eq-tau1}
\end{align}
Figure~\ref{fig-LossDelay}(b) shows the variation of the output pulse delay depending on the input states. 
These variations cause degradation in the coherence of the photonic qubits, due to a temporal mode-mismatch error~\cite{rohde2005optimal}.
The gate fidelity reduction due to the pulse delay variation (pulse-delay error) is cancelled by matching the three pulse delays $\tau_{0}, \tau_{1}$, and $\tau_{\rm{ref}}$, where $\tau_{\rm{ref}}$ is the pulse delay of the vertical polarization states from the external mirror. 
We can minimize the pulse-delay error by minimizing $|\tau_0-\tau_1|$, as the delay of the pulse from the external mirror $\tau_{\rm{ref}}$ can be freely shifted by adjusting the position of the external mirror. Then, $\tau_{\rm{ref}}$ is set to $\tau_{\rm{ref}}=(\tau_{0}+\tau_{1})/2$ to minimize the pulse-delay error as shown in Fig~\ref{fig-LossDelay}(b).

\begin{figure}[tb]
\includegraphics[clip,width=60mm]{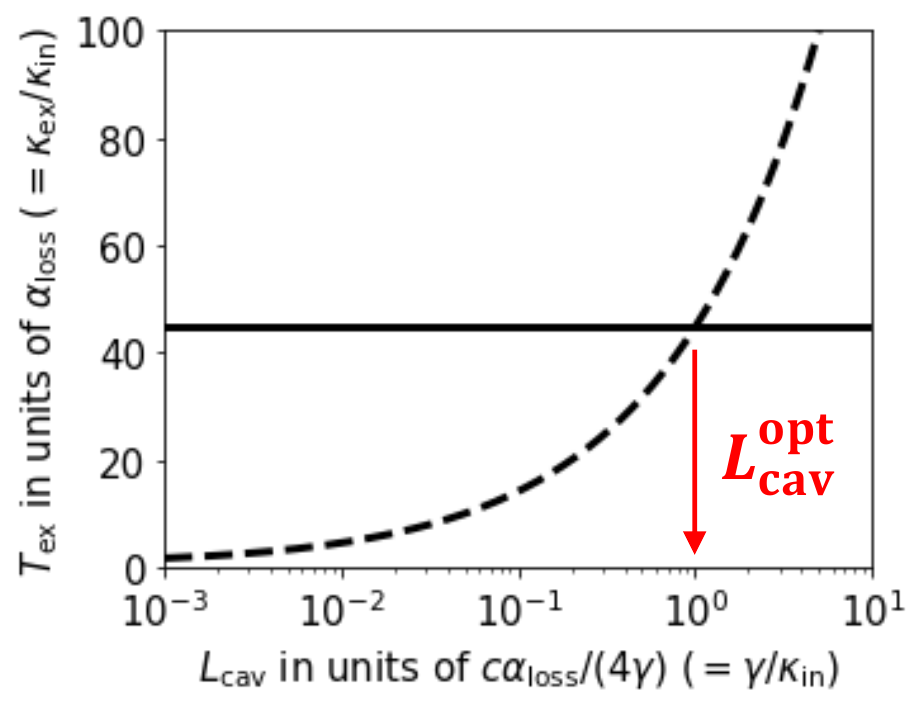}
\caption{Transmittance minimizing the photon loss probability $T_{\rm{ex}}=T_{\rm{ex}}^{\rm{loss}}$  in Eq.\,(\ref{eq-Texloss}) (solid line) and the transmittance minimizing the pulse-delay error rate $T_{\rm{ex}}=T_{\rm{ex}}^{\rm{delay}}$ in Eq.\,(\ref{eq-Texdelay}) (dashed curve), as a function of the cavity length $L_{\rm{cav}}$.
They intersect at optimal cavity length $L_{\rm{cav}}=L_{\rm{cav}}^{\rm{opt}}$. These plots are calculated by setting the internal cooperativity to $C_{\rm{in}}=1000$. We assume that $\alpha_{\rm{loss}}$ is independent of $L_{\rm{cav}}$.}
\label{fig-Lcavopt}
\end{figure}

Based on the formulation of error sources presented above, we derive optimal cavity parameters for the CPF gate.
It is known that a measure of the norm reduction $l_{0}+l_{1}$ and a measure of the loss imbalance $|l_{0}-l_{1}|$ can be minimized simultaneously by setting the external cavity field decay rate as $\kappa
_{\rm{ex}}=\kappa_{\rm{in}}\sqrt{2C_{\rm{in}}+1}$~\cite{goto2010condition}.
This condition is rewritten using Eqs.~(\ref{eq-g})-(\ref{eq-kappa-ex}) as
\begin{align}
	T_{\rm{ex}}=T^{\rm{loss}}_{\rm{ex}}\equiv \alpha_{\rm{loss}}\sqrt{\frac{2}{\tilde{A}_{\rm{eff}}\alpha_{\rm{loss}}}+1}\approx\sqrt{\frac{2\alpha_{\rm{loss}}}{\tilde{A}_{\rm{eff}}}}.
	\label{eq-Texloss}
\end{align}
For the final approximation, we assume ${2C_{\rm{in}}\gg1}$, or ${2/(\tilde{A}_{\rm{eff}}\alpha_{\rm{loss}})\gg1}$ (this approximation is also applied to the following equations). 
On the other hand, from Eqs. (\ref{eq-tau0}) and (\ref{eq-tau1}), the condition $\tau_{0}=\tau_{1}$, namely the condition minimizing a measure of the pulse-delay error $|\tau_{0}-\tau_{1}|$, is given by
\begin{align}
	T_{\rm{ex}}=T^{\rm{delay}}_{\rm{ex}}&\equiv\sqrt{\frac{8L_{\rm{cav}}\gamma}{c}\left(\alpha_{\rm{loss}}+\frac{1}{\tilde{A}_{\rm{eff}}}\right) + \alpha_{\rm{loss}}^{2}}\notag\\
	&\approx\sqrt{\frac{8L_{\rm{cav}}\gamma}{c\tilde{A}_{\rm{eff}}}}.
	\label{eq-Texdelay}
\end{align}
Equation~(\ref{eq-Texdelay}) is equivalent to $g/\kappa\approx1$.
Figure~\ref{fig-Lcavopt} shows $T_{\rm{ex}}^{\rm{loss}}$ and $T_{\rm{ex}}^{\rm{delay}}$  as a function of $L_{\rm{cav}}$.
As mentioned above, on the curve of $T_{\rm{ex}}=T_{\rm{ex}}^{\rm{delay}}$ (dashed curve), the pulse delay causes no error because $|\tau_0-\tau_1|=0$. On the line $T_{\rm{ex}}=T_{\rm{ex}}^{\rm{loss}}$, $l_0+l_1$ and $|l_0-l_1|$ are constant as seen in Eqs.\,(\ref{eq-l0}) and (\ref{eq-l1}). These facts indicate that the total error probability in the CPF gate is minimized at the intersection of the functions, namely $T_{\rm{ex}}^{\rm{loss}} = T_{\rm{ex}}^{\rm{delay}}$.
From Eqs.~(\ref{eq-Texloss}) and (\ref{eq-Texdelay}), we obtain the optimal cavity length 
\begin{equation}
	L_{\rm{cav}}^{\rm{opt}}=\frac{c}{4\gamma(\tilde{A}_{\rm{eff}}+\alpha_{\rm{loss}}^{-1})}\approx\frac{c\alpha_{\rm{loss}}}{4\gamma}.
	\label{eq-Lopt}
\end{equation}
Note that Eq.~(\ref{eq-Lopt}) can be rewritten as $\kappa_{\rm{in}} \approx \gamma$. 
In other words, adjusting the cavity length to balance the internal-loss rate and the atomic polarization decay rate minimizes the total error probability.

From Eq.~(\ref{eq-basic2}) and the balancing condition $\kappa_{\rm{in}}/\gamma=1$, we establish some guidelines towards optimal cavity design for the CPF gate.
For a typical free-space cavity, e.g., $\kappa_{\rm{in}}/\gamma = 0.067$~\cite{daiss2021quantum}, it is desirable to increase $\kappa_{\rm{in}}$. 
However, considering the relation $\kappa_{\rm{in}}\propto \alpha_{\rm{loss}}/L_{\rm{cav}}$, it is challenging to further decrease $L_{\rm{cav}}$ or increase $\alpha_{\rm{loss}}$ with decreasing $\tilde{A}_{\rm{eff}}$ to maintain $C_{\rm{in}}$ when using a conventional bulk mirror.
In contrast, nanofiber cavities with, e.g., $\kappa_{\rm{in}}/\gamma=0.04$~\cite{ruddell2020ultra,asaoka2021requirements}, have small $\tilde{A}_{\rm{eff}}$ to the limit, hence shortening $L_{\rm{cav}}$ may be the only strategy.
For other cavities with large $\kappa_{\rm{in}}/\gamma$, such as fiber cavities ($\kappa_{\rm{in}}/\gamma=8.3$~\cite{macha2020nonadiabatic}) or nanophotonic cavities ($\kappa_{\rm{in}}/\gamma=460$~\cite{samutpraphoot2020strong}), $\alpha_{\rm{loss}}/L_{\rm{cav}}$ should be reduced.

In the above discussion, we have assumed that the dependence of $\tilde{A}_{\rm{eff}}$ and $\alpha_{\rm{loss}}$ on the cavity length is negligible. 
This is justified in typical optical cavities, where the bulk loss of photons (due to absorption and scattering in the cavity medium) is negligible.
On the other hand, in some systems, including photonic crystal cavities, the bulk loss cannot be neglected~\cite{vahala2003optical,kavokin2017microcavities,chang2018colloquium}.
Due to the bulk loss, the optimal cavity length that minimizes the respective photon loss probability and pulse-delay error differs. Therefore, the optimal cavity length is not uniquely determined and varies depending on the application.
We discuss the bulk-loss effect in detail in Supplemental Material III~\cite{sup}.

\begin{figure*}[tb]
\includegraphics[clip,width=160mm]{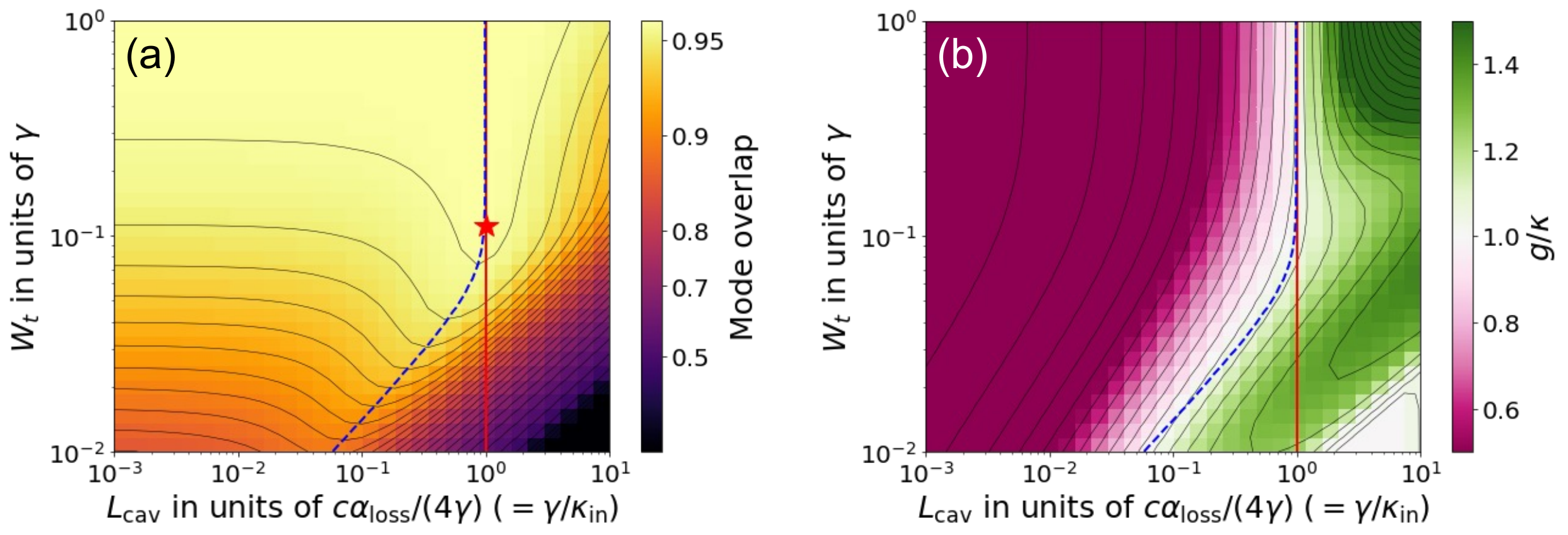}
\caption{(a) Simulation result of the mode overlap as a function of $W_{t}$ and $L_{\rm{cav}}$ when $T_{\rm{ex}}$ is optimized numerically. 
The red solid line represents $L_{\rm{cav}}=L_{\rm{cav}}^{\rm{opt}}$ in Eq.~(\ref{eq-Lopt}), and the blue dashed curve represents the cavity length optimized numerically for each pulse duration. The red star represents five times the right-hand side of Eq.~(\ref{eq-critical}), that is, $W_{t}=5/(\gamma\sqrt{2C_{\rm{in}}})$.
The pulse delay of the photon reflecting the external mirror is set to $\tau_{\rm{ref}}=(\tau_{0}+\tau_{1})/2$, and the internal cooperativity is set to $C_{\rm{in}}=1000$.
(b) The coupling rate in the same simulation of (a). The optimized cavity length (blue dashed curve) and $g/\kappa=1$ region are in good agreement for any pulse duration.}
\label{fig-result}
\end{figure*}

Here, we discuss the shift of the optimal cavity length due to the effect of the pulse distortion represented by the higher order terms in Eqs.~(\ref{eq-L0approx2}) and (\ref{eq-L1approx2}) by numerically solving Eqs.~(S.1)-(S.5) in Supplemental Material I~\cite{sup}.
We consider an input photon wave packet having a Gaussian shape as ${{f^{\rm{in}}(t)=\sqrt{\frac{1}{\sqrt{\pi}W_{\rm{t}}}}\exp\left(-\frac{t^{2}}{2W_{\rm{t}}^{2}}\right)}}$, where  $W_{\rm{t}}$ represents the pulse duration~\cite{memo}.
To estimate the total error rate of the CPF gate operation, we calculate {mode overlap defined by $|\langle\Psi_{\rm{id}}|\Psi_{\rm{out}}\rangle|^{2}$, where $|\Psi_{\rm{out}}\rangle$ and $|\Psi_{\rm{id}}\rangle$ are an output state of the non-Hermitian Hamiltonian time evolution and an ideal output state, respectively.
We assume that the input state is the equal superposition state, which is the worst fidelity estimation of all input states.
This is because our error model only includes phase error other than photon loss, and the equal superposition state is most sensitive to the phase error.
This mode overlap is also consistent with the entanglement fidelity~\cite{nielsen2002} in the case of CPF gate} (see Supplemental Material VI~\cite{sup}).

Figure~\ref{fig-result}(a) shows the mode overlap as a function of the Gaussian pulse duration $W_{t}$ and the cavity length $L_{\rm{cav}}$. 
The red solid line denotes the cavity length in Eq.\,(\ref{eq-Lopt}), and the blue dashed curve denotes the cavity length maximizing the mode overlap calculated numerically for each pulse duration. They are in very good agreement for $W_{\rm{t}} \gtrsim 0.1\gamma^{-1}$.

For $W_{\rm{t}} \lesssim 0.1\gamma^{-1}$, on the other hand, there is a discrepancy between the red solid line and blue dashed curve due to the effect of the higher-order terms of $\Delta$ in Eqs.~(\ref{eq-L0approx2}) and (\ref{eq-L1approx2}).
The condition of the pulse duration where the higher order terms can be considered negligible, that is, the pulse duration at the branch of the red solid line and the blue dashed curve in Fig.~\ref{fig-result}(a), can be roughly estimated from the point where the values of the first and second terms in Eqs.~(\ref{eq-L0approx2}) and (\ref{eq-L1approx2}) are equal.
From the discussion in Supplemental Material IV~\cite{sup}, this condition can be written as ${W_{\rm{t}}\gg \max(1/\kappa,\kappa/g^{2})}$ when the cavity length is optimal as shown in Eq.~(\ref{eq-Lopt}). Then, using Eqs.~(\ref{eq-basic1}) and (\ref{eq-Texloss}), we obtain the pulse duration condition where the first-order approximation is valid as 
\begin{equation}
	W_{\rm{t}}\gg \max\left(\frac{1}{\kappa},\frac{\kappa}{g^2}\right)\approx \frac{1}{\gamma\sqrt{2C_{\rm{in}}}}.
	\label{eq-critical}
\end{equation}
When the second-order or higher terms are raised, the mode overlap remarkably drops. Then, the pulse duration should be at least as long as the right-hand side of Eq.~(\ref{eq-critical}), indicated by the red star in Fig.~\ref{fig-result}(a).
We note that the speed of the CPF gate is characterized by the right-hand side of Eq.~(\ref{eq-critical}), determined by $C_{\rm{in}}$ and $\gamma$ under the optimal cavity condition.  

Figure~\ref {fig-result}(b) shows the result of the coupling rate $g/\kappa$ for the same simulation as Fig~\ref {fig-result}(a). From Fig.~\ref{fig-result}(b), we can obtain a hint regarding the optimal cavity design even when the first-order approximation is invalid. 
The fact that the numerically optimized $L_{\rm{cav}}$ (blue dashed curve) is in good agreement with $g/\kappa=1$ (white region) suggests that the condition for minimizing the temporal mode-mismatch error, i.e., $\tau_{0}=\tau_{1}$ or Eq.~(\ref{eq-Texdelay}), holds even for $W_{\rm{t}} \lesssim 0.1\gamma^{-1}$. 
In other words, $T_{\rm{ex}}^{\rm{loss}}$ in Eq.~(\ref{eq-Texloss}) is no longer valid, even though $T_{\rm{ex}}^{\rm{delay}}$ in Eq.~(\ref{eq-Texdelay}) is still valid in this region.
Thus, the optimal cavity length becomes shorter than that in Eq.~(\ref{eq-Lopt}) for such short pulses. 
This argument is supported by the discussion in Supplemental Material V~\cite{sup}, where we show that the higher-order term effect is corrected by decreasing $L_{\rm{cav}}$. 

We have investigated the optimal cavity parameters for CQED-based CPF gates with photons having finite pulse duration. 
From the response function of the CQED system, we formulated two error sources: the photon loss probability and the error rate due to temporal mode-mismatch of the output photon pulse, which we take into account as pulse delay.
We then derived the optimal condition of the tunable cavity parameters. 
As a result, the balance of the photonic and atomic decay rate ($\kappa_{\rm{in}}=\gamma$) was derived for optimization of the cavity length.
This means that there is an optimal cavity length for finite pulse duration, whereas the error sources do not limit cavity length for the long-pulse limit. 
Furthermore, we performed a numerical simulation of the dynamics of the CQED system for fast gate operations where temporal mode-mismatch errors beyond the first-order approximation are not negligible.
However, in such a regime, error probabilities in the CPF gate are quite high (over five percent), indicating that our optimal cavity design obtained analytically is basically valid for the purpose of scalable quantum computing.
Besides, the analytical treatment of temporal mode-mismatch we have constructed will be applicable for general CQED systems in which photons are exchanged in and out of a cavity, including the quantum memory~\cite{reiserer2015cavity}, quantum network construction~\cite{long4}, and quantum cluster state generation~\cite{long10}.

Regarding future work, a similar argument should be valid for other CQED-based entangling gates~\cite{reiserer2015cavity}.
We note that the optimal cavity parameters and pulse duration condition for CPF gates derived in this work are consistent for single-photon generation~\cite{utsugi2022gaussian}. Thus, it is possible to realize both single-photon generation and the CPF gate using the same CQED setup, which appears to be efficient for the construction of CQED-based quantum computing systems.

\section*{Acknowledgement}
\noindent The authors thank Hayato Goto, Rina Kanamoto, Samuel Ruddell, Karen Webb, and Akinori Suenaga for their helpful comments. 
This work was supported by JST CREST, Grant Number. JPMJCR1771, and JST Moonshot R$\&$D, Grant Numbers JPMJMS2061, and JPMJMS2268, Japan.

\newpage
\onecolumngrid

%
%

\section*{Supplemental Material}

{
\section{\label{apI}Derivation of $L_{0}(\Delta)$ and $L_{1}(\Delta)$}
}
Introducing the internal cavity field decay rate $\kappa_{\rm{in}}$, the total cavity field decay rate ${\kappa=\kappa_{\rm{in}}+\kappa_{\rm{ex}}}$, and the atomic polarization decay rate $\gamma$, we obtain the equations of motion by using Eqs.~(3) and (4) in the main text as~(see Refs.~[11,12] in the main text).
{\begin{align}
	&\dot{B_{0}}(t)=-\kappa_{\rm{in}} B_{0}(t)+\sqrt{\frac{\kappa_{\rm{ex}}}{\pi}} \int^{\infty}_{-\infty}d\omega F_{0}(t,\omega),\label{A1}\\
	&\dot{B_{1}}(t)=gB_{e}(t)-\kappa_{\rm{in}} B_{1}(t)+\sqrt{\frac{\kappa_{\rm{ex}}}{\pi}} \int^{\infty}_{-\infty}d\omega F_{1}(t,\omega),\label{A2}\\
	&\dot{B_{e}}(t)=-gB_{1}(t)-\gamma B_{e}(t),\label{A3}\\
	&\dot{F_{0}}(t,\omega)=-i(\omega-\omega_{c})F_{0}(t,\omega)-\sqrt{\frac{\kappa_{\rm{ex}}}{\pi}}B_{0}(t),\label{A4}\\ 
	&\dot{F_{1}}(t,\omega)=-i(\omega-\omega_{c})F_{1}(t,\omega)-\sqrt{\frac{\kappa_{\rm{ex}}}{\pi}}B_{1}(t),\label{A5}
\end{align}}

\noindent
where
{\begin{align}
	&B_{0}(t)=e^{i(\omega_{\rm{c}}-\omega_{01})t}\beta_{0}(t),\\ 
	&B_{1}(t)=e^{i\omega_{\rm{c}}t}\beta_{1}(t),\\ 
	&B_{e}(t)=e^{i\omega_{\rm{c}}t}\beta_{e}(t),\\ 
	&F_{0}(t,\omega)=e^{i(\omega_{\rm{c}}-\omega_{01})t}f_{0}(t,\omega),\\
	&F_{1}(t,\omega)=e^{i\omega_{\rm{c}}t}f_{1}(t,\omega),
\end{align}}
\noindent
By solving these differential equations under input and output conditions, we derive
\begin{align}
(\rm{\ref{A1}, \ref{A4}}) &\rightarrow 
	\begin{cases}
		{\dot{B_{0}}(t) = -(\kappa_{\rm{ex}}+\kappa_{\rm{in}}) B_{0}(t)+\sqrt{\frac{\kappa_{\rm{ex}}}{\pi}}\int_{-\infty}^{\infty} d\omega e^{-i\Delta_{c} t}F_{0}(0,\omega),}\\
		{\dot{B_{0}}(t) = (\kappa_{\rm{ex}}-\kappa_{\rm{in}})  B_{0}(t)+\sqrt{\frac{\kappa_{\rm{ex}}}{\pi}}\int_{-\infty}^{\infty} d\omega e^{-i\Delta_{c} (t-T)}F_{0}(T,\omega),}
	\end{cases}\label{A015}\\
(\rm{\ref{A2}, \ref{A5}}) &\rightarrow 
	\begin{cases}
		{\dot{B_{1}}(t) = gB_{e}(t)-(\kappa_{\rm{ex}}+\kappa_{\rm{in}}) B_{1}(t)+\sqrt{\frac{\kappa_{\rm{ex}}}{\pi}}\int_{-\infty}^{\infty} d\omega e^{-i\Delta_{c} t}F_{1}(0,\omega),}\\
		{\dot{B_{1}}(t) = gB_{e}(t)+(\kappa_{\rm{ex}}-\kappa_{\rm{in}}) B_{1}(t)+\sqrt{\frac{\kappa_{\rm{ex}}}{\pi}}\int_{-\infty}^{\infty} d\omega e^{-i\Delta_{c} (t-T)}F_{1}(T,\omega),}
	\end{cases}\label{A017}\\
(\rm{\ref{A3}}) &\rightarrow \dot{B_{e}}(t) = -gB_{g}(t) -\gamma B_{e}(t),\label{A019}
\end{align}
where $T$ denotes the end time of the CPF gate and $\Delta_{c}\equiv\omega-\omega_{c}$.
In deriving above equations, we used $\int_{0}^{t}\beta_{m}(t')\delta(t-t')dt'=\beta_{m}(t)/2~(m\in \{0,1,e\})$. 
Then, we obtain the input-output relation in the rotation frame as
\begin{align}
-2\kappa_{\rm{ex}} B_{0}(t)+\sqrt{\frac{\kappa_{\rm{ex}}}{\pi}}\int_{-\infty}^{\infty} d\omega e^{-i\Delta_{c} t}F_{0}(0,\omega) = \sqrt{\frac{\kappa_{\rm{ex}}}{\pi}}\int_{-\infty}^{\infty} d\omega e^{-i\Delta_{c} (t-T)}F_{0}(T,\omega),\label{A020}\\
-2\kappa_{\rm{ex}} B_{1}(t)+\sqrt{\frac{\kappa_{\rm{ex}}}{\pi}}\int_{-\infty}^{\infty} d\omega e^{-i\Delta_{c} t}F_{1}(0,\omega) = \sqrt{\frac{\kappa_{\rm{ex}}}{\pi}}\int_{-\infty}^{\infty} d\omega e^{-i\Delta_{c} (t-T)}F_{1}(T,\omega).\label{A021}
\end{align}
By inverse Fourier transform of Eqs~(\ref{A015})-(\ref{A021}) with $\tilde{B_{0}}(\omega)=\frac{1}{\sqrt{2\pi}}\int_{-\infty}^{\infty} dt B_{0}(t)e^{i\Delta_{c}t}$, $\tilde{B_{1}}(\omega)=\frac{1}{\sqrt{2\pi}}\int_{-\infty}^{\infty} dt B_{1}(t)e^{i\Delta_{c}t}$, and $\tilde{B_{e}}(\omega)=\frac{1}{\sqrt{2\pi}}\int_{-\infty}^{\infty} dt B_{e}(t)e^{i\Delta_{c}t}$, we obtain
\begin{align}
-i\Delta_{c} \tilde{B_{0}}(\omega) &= -\kappa \tilde{B_{0}}(\omega) + \sqrt{2\kappa_{\rm{ex}}}F_{0}(0,\omega),\\
-i\Delta_{c} \tilde{B_{1}}(\omega) &= g\tilde{B_{e}}(\omega) -\kappa \tilde{B_{1}}(\omega) + \sqrt{2\kappa_{\rm{ex}}}F_{1}(0,\omega),\\
-i\Delta_{c} \tilde{B_{e}}(\omega) &= -g\tilde{B_{1}}(\omega) - \gamma \tilde{B_{e}}(\omega),\\
e^{i\Delta_{c} T}F_{0}(T,\omega) &= F_{0}(0,\omega) - \sqrt{2\kappa_{\rm{ex}}}\tilde{B_{0}}(\omega),\\
e^{i\Delta_{c} T}F_{1}(T,\omega) &= F_{1}(0,\omega) - \sqrt{2\kappa_{\rm{ex}}}\tilde{B_{1}}(\omega).
\end{align}
From these equations, we obtain $L_{0}(\Delta)$ and $L_{1}(\Delta)$ in Eqs.~(5) and (6) in the main text as
\begin{align}
e^{i\Delta_{c} T}F_{0}(T,\omega) &= -\frac{\kappa_{\rm{ex}}-\kappa_{\rm{in}}+i\Delta_{c}}{\kappa_{\rm{ex}}+\kappa_{\rm{in}}-i\Delta_{c}}F_{0}(0,\omega)=L_{0}(\Delta)F_{0}(0,\omega),\\
e^{i\Delta_{c} T}F_{1}(T,\omega) &= -\frac{\kappa_{\rm{ex}}-\kappa_{\rm{in}}+i\Delta_{c}-\frac{g^{2}}{\gamma-i\Delta_{c}}}{\kappa_{\rm{ex}}+\kappa-i\Delta_{c}+\frac{g^{2}}{\gamma-i\Delta_{c}}}F_{1}(0,\omega)=L_{1}(\Delta)F_{1}(0,\omega),
\end{align}
where $\Delta $ in the main text is defined by $\Delta = -\Delta_{c}$.
\\
\\

\section{\label{apII}Derivation of pulse delay}
Here, we show that $l_{n}$ and $\tau_{n}$ ($n\in\{0,1\}$) defined by Eqs.~(16)-(19) in the main text represent the photon loss probabilities and the pulse delays, respectively.
The functions $L_{n}(\Delta)~(n\in \{0,1\})$ in Eqs.~(5) and (6) in the main text are expanded by 
\begin{equation}
{L_{n}(\Delta)=L_{n}(0)+\left.\frac{dL_{n}(\Delta)}{d\Delta}\right|_{\Delta=0}\Delta+O(\Delta^{2})}.
\label{eq-b1}
\end{equation}
When $\Delta$ is sufficiently small, this becomes
\begin{equation}
L_{n}(\Delta) \approx L_{n}(0)\exp\left[\left.\frac{1}{L_{n}(0)}\frac{dL_{n}(\Delta)}{d\Delta}\right|_{\Delta=0}\Delta\right].
\label{eq-b2}
\end{equation}
Comparing Eq.~(\ref{eq-b1}) and Eqs.~(14) and (15) in the main text, we find that the first term of Eq.~(\ref{eq-b1}) is real ($L_{n}(0)\in\mathbb{R}$) and the second term of Eq.~(\ref{eq-b1}) is imaginary ($\left.\frac{dL_{n}(\Delta)}{d\Delta}\right|_{\Delta=0}\in\mathbb{I}$), thus, the inside of the exponential function in Eq.~(\ref{eq-b2}) is imaginary.
Then, Eq.~(\ref{eq-b2}) is rewritten by
\begin{equation}
L_{n}(\Delta)\approx L_{n}(0)\exp(-i\Delta \tau_{n}).
\label{eq-b3}
\end{equation}
where we used the relation ${-i\tau_{n}=\left.\frac{1}{L_{n}(0)}\frac{dL_{n}(\Delta)}{d\Delta}\right|_{\Delta=0}} (n\in\{0,1\})$ introduced by Eqs.~(18) and (19) in the main text.
From the time shifting property of a Fourier transform $\mathbf{F}$ on a pulse waveform $f_{n}(t)~(n\in\{0,1\})$,
\begin{equation}
\mathbf{F} [ f_{n}(t-\tau_{n}) ](\Delta) = \mathbf{F} [ f_{n}(t) ] \exp(-i \Delta \tau_{n}),
\label{eq-b4}
\end{equation}
we obtain the output pulse waveform $f^{\rm{out}}_{n}(t)~(n\in\{0,1\})$ as 
\begin{align}
f^{\rm{out}}_{n}(t) 
&= 
\mathbf{F}^{-1}\left\{\mathbf{F}[f_{n}(t)](\Delta)L_{n}(\Delta)\right\}\notag\\
&=
L_{n}(0)\mathbf{F}^{-1}\left\{\mathbf{F}[f_{n}(t)](\Delta)\exp(-i\Delta\tau_{n})\right\}\notag\\
&=L_{n}(0)f(t-\tau_{n}).
\label{eq-b5}
\end{align}
Equation~(\ref{eq-b5}) shows that the output pulse is multiplied by $L_{n}(0)$, i.e., the photon loss probabilities can be represented by $l_{n}\equiv1-L_{n}(0)^{2}$ and the output pulse is delayed to $\tau_{n}$.

\section{\label{apIII}Effect of bulk loss}
The roundtrip loss in a cavity with non-negligible bulk loss is typically written as $\alpha_{\rm{loss}}=1-\exp\left[-(\alpha'+\beta L_{\rm{cav}})\right]\approx \alpha'+\beta L_{\rm{cav}}$, where the approximation holds when $\alpha'+\beta L_{\rm{cav}}\ll1$. 
The first term is mainly due to scattering by the cavity mirrors, and the second term is mainly due to bulk loss.

\begin{figure}[h]
\includegraphics[width=0.5 \textwidth]{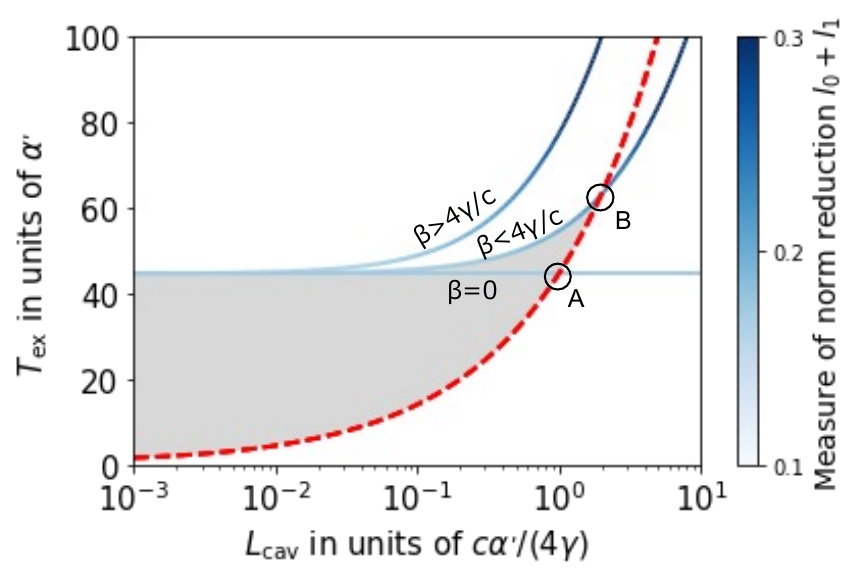}
\caption{Transmittance minimizing the photon-loss error $T_{\rm{ex}}=T_{\rm{ex}}^{\rm{loss}}$ (blue gradation solid curves) and the transmittance minimizing the pulse-delay error $T_{\rm{ex}}=T_{\rm{ex}}^{\rm{delay}}$ (red dashed curve), as a function of the cavity length $L_{\rm{cav}}$.
$T_{\rm{ex}}^{\rm{loss}}$ is plotted for the three cases where $\beta=0$, $0<\beta<4\gamma/c$, and $4\gamma/c<\beta$.
These curves are calculated by setting $C_{\rm{in}}=1000$ when $\beta=0$.
For $\beta>0$, the measure of the norm reduction $l_{0}+l_{1}$ depends on the cavity length as represented by the color of the curves.
The shaded area indicates suitable parameter sets ($T_{\rm{ex}}$, $L_{\rm{cav}}$) in the case of $0<\beta<4\gamma/c$, depending on the application.
}
\label{fig-bulk}
\end{figure}

Figure~\ref{fig-bulk} shows $T_{\rm{ex}}^{\rm{loss}}$ and $T_{\rm{ex}}^{\rm{delay}}$ in the main text as a function of $L_{\rm{cav}}$ for ${\alpha_{\rm{loss}}=\alpha'+\beta L_{\rm{cav}}}$.
For the $\beta=0$ case, as shown in the main text, any type of error is minimized for $L_{\rm{cav}}=L_{\rm{cav}}^{\rm{opt}}$ in Eq.~(22) in the main text, i.e., at the intersection A in Fig.~\ref{fig-bulk}.
For the $\beta>0$ case, considering non-negligible bulk loss, the pulse-delay error and the error due to the unwanted state-vector rotation are zero when $T_{\rm{ex}}=T_{\rm{ex}}^{\rm{delay}}$ and $T_{\rm{ex}}=T_{\rm{ex}}^{\rm{loss}}$, respectively, for any cavity length as in the $\beta=0$ case. However, the measure of the norm reduction $l_{0}+l_{1}$ increases with $L_{\rm{cav}}$ when $T_{\rm{ex}}=T_{\rm{ex}}^{\rm{loss}}$.
Therefore, there is no longer a cavity length that simultaneously minimizes all types of errors.
Thus we can only conclude the following: An optimal cavity-parameter set ($T_{\rm{ex}}, L_{\rm{cav}}$) for each application is always somewhere in the shaded area in Fig.~\ref{fig-bulk}. In other words, if a cavity-parameter set outside of the shaded area was chosen, a more suitable set within the shaded area for which at least one of the errors is smaller can be found.
The cavity length of the intersection B is derived as
\begin{equation}
	L_{\rm{cav}}\approx\frac{c\alpha'}{4\gamma-c\beta}
	\label{eq-Lopt2}
\end{equation}
from Eq. (22) in the main text. For $\beta>4\gamma/c$, the intersection no longer exists, and finding an optimal cavity-parameter set is less straightforward.
Even in such a case, the optimal $T_{\rm{ex}}$ can be found in the range of $T_{\rm{ex}}^{\rm{delay}} \leq T_{\rm{ex}} \leq T_{\rm{ex}}^{\rm{loss}}$. 

\section{\label{apIV}Condition for the first order approximation}
We expand Eq.~(6) in the main text by $\Delta$ as
\begin{align}
	L_{1}(\Delta) &= \left(1-\frac{2\gamma\kappa_{\rm{ex}}}{g^{2}+\gamma\kappa}\right) - i \frac{2\kappa_{\rm{ex}}(g^{2}-\gamma^{2})}{(g^{2}+\gamma\kappa)^{2}}\Delta -\frac{2\kappa_{\rm{ex}}(2g^{2}\gamma+g^{2}\kappa-\gamma^{3})}{(g^{2}+\gamma\kappa)^{3}}\Delta^{2}+O(\Delta^{3}).
	\label{eq-L1approxA}
\end{align}
Thus, the condition that the higher order terms above second order are negligible is
\begin{gather}
	\left|\frac{2\kappa_{\rm{ex}}(g^{2}-\gamma^{2})}{(g^{2}+\gamma\kappa)^{2}}\right|\Delta\gg
	\left|\frac{2\kappa_{\rm{ex}}(2g^{2}\gamma+g^{2}\kappa-\gamma^{3})}{(g^{2}+\gamma\kappa)^{3}}\right|\Delta^{2}\\
	\rightarrow
	\Delta\ll\frac{(g^{2}-\gamma^{2})(g^{2}+\gamma\kappa)}{2g^{2}\gamma+g^{2}\kappa-\gamma^{3}}\approx\frac{g^{2}}{\kappa},\label{negl}
\end{gather}
where the approximation in the second line holds when $g\approx\kappa\gg\gamma\approx\kappa_{\rm{in}}$ [the optimal condition as shown in Eqs.~(12), (21), and (22) in the main text] is satisfied.

When $g=0$, we obtain the condition for $L_{0}(\Delta)$ as
\begin{gather}
	\Delta\ll\kappa.\label{neg0}
\end{gather}
Since the condition $\Delta \ll \delta$ means that the pulse duration $W_{\rm{t}}$ is required to be sufficiently longer than $1/\delta$, we finally obtain the condition on the pulse duration for the first order approximation, ${W_{\rm{t}}\gg \max(1/\kappa,\kappa/g^{2})}$.

\section{\label{apV}Effect of higher-order terms}
\begin{figure}[h]
\includegraphics[width=0.4 \textwidth]{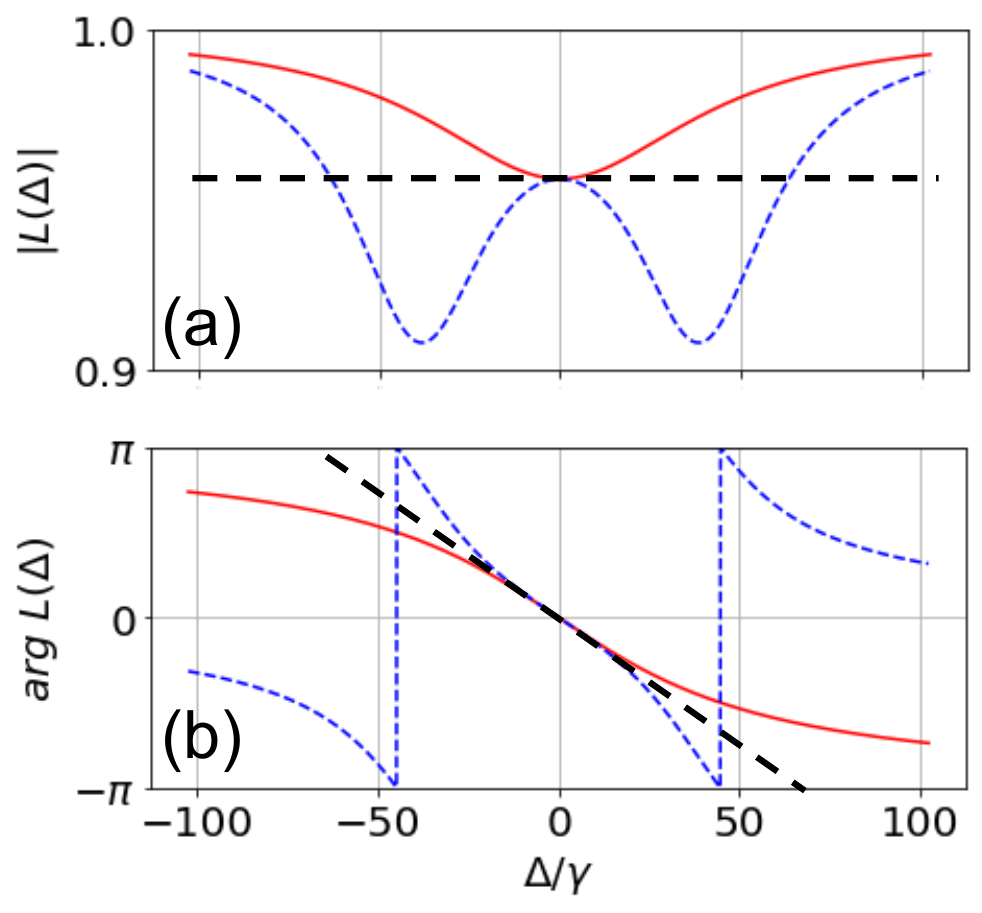}
\caption{The plot of amplitude reflection coefficients: (a) amplitudes and (b) phases of $L_{0}(\Delta)$ (red solid curve) and $L_{1}(\Delta)$ (blue dashed curve) in the case of $\kappa_{\rm{in}}=\gamma$ and $T_{\rm{ex}}=T_{\rm{ex}}^{\rm{loss}}=T_{\rm{ex}}^{\rm{delay}}$.
The black dashed lines represent a first-order approximation of amplitude and phase. 
The phase of $L_{0}(\Delta)$ is shifted by $\pi$ from the original value. The system parameters are set to $(\kappa_{\rm{in}}/\gamma,C_{\rm{in}})=(1,1000)$.}
\label{fig-highorder}
\end{figure}

We show that the higher-order term effect is corrected by decreasing $L_{\rm{cav}}$.
First, we consider the optimization of $T_{\rm{ex}}$ for a certain cavity length.
When the pulse duration is so short that the second-order term of $\Delta$ is not negligible, as shown in Fig.~\ref{fig-highorder}, $l_{0}(\Delta)=1-|L_{0}(\Delta)|^{2}$ decreases and $l_{1}(\Delta)=1-|L_{1}(\Delta)|^{2}$ increases as $|\Delta|$ increases. On the other hand, $\tau_{0}(\Delta)=|\frac{d}{d\Delta}\rm{arg}L_{0}(\Delta)|$ decreases and $\tau_{1}(\Delta)=|\frac{d}{d\Delta}\rm{arg}L_{1}(\Delta)|$ increases as $|\Delta|$ increases. 
Therefore, $T_{\rm{ex}}$ needs to be optimized to compensate for this effect.
The first derivatives of the photon loss probabilities and the pulse delays for $T_{\rm{ex}}$ are given by
{ \begin{align}
\frac{dl_{0}}{dT_{\rm{ex}}} &= - \frac{4 \alpha_{\rm{loss}} \left(T_{\rm{ex}} - \alpha_{\rm{loss}}\right)}{\left(T_{\rm{ex}} + a_{\rm{loss}}\right)^{3}}<0,\\
\frac{dl_{1}}{dT_{\rm{ex}}} &= \frac{4 \left( 2/\tilde{A}_{\rm{eff}}+\alpha_{\rm{loss}}\right) \left(2/\tilde{A}_{\rm{eff}} - T_{\rm{ex}} + \alpha_{\rm{loss}} \right)}{\left(2/\tilde{A}_{\rm{eff}} + T_{\rm{ex}} + \alpha_{\rm{loss}} \right)^{3}}>0,\\
\frac{d\tau_{0}}{dT_{\rm{ex}}} &= - \frac{8 L_{\rm{cav}} \left(T_{\rm{ex}}^{2} + \alpha_{\rm{loss}}^{2}\right)}{c \left(T_{\rm{ex}}^{2} - \alpha_{\rm{loss}}^{2}\right)^{2}}<0,\\
\frac{d\tau_{1}}{dT_{\rm{ex}}} &= \frac{4 \left(c/\tilde{A}_{\rm{eff}}- 2 L_{\rm{cav}} \gamma \right) \left( T_{\rm{ex}}^{2} + \alpha_{\rm{loss}}^{2} + 4 \alpha_{\rm{loss}}/\tilde{A}_{\rm{eff}} + 4/\tilde{A}_{\rm{eff}}^{2}\right)}{c \gamma \left( T_{\rm{ex}} - \alpha_{\rm{loss}} - 2/\tilde{A}_{\rm{eff}}\right)^{2} \left(T_{\rm{ex}} + \alpha_{\rm{loss}} + 2/\tilde{A}_{\rm{eff}} \right)^{2}}>0,
\label{D4}
\end{align}}
where we used the relation of Eq.~(13) in the main text at each inequality and $L_{\rm{cav}}\approx L_{\rm{cav}}^{\rm{opt}}\approx c\alpha_{\rm{loss}}/(4\gamma)$ in the inequality of Eq.~(\ref{D4}).
From these equations, it is expected that a smaller value of $T_{\rm ex}$ than that obtained using the first-order approximation is optimal for the CPF gate when considering the higher-order terms.
Thus, the optimal cavity length becomes shorter than that in Eq.~(22) in the main text because $T_{\rm ex}\propto\sqrt{L_{\rm cav}}$ [see Eq. (21)]. Thus for short pulses, and therefore for fast gate operations, the temporal mode-mismatch errors beyond the first-order approximation are not negligible.

\section{\label{apVI}Relation between the mode overlap and the entanglement fidelity in the CPF gate}

{We used the ''mode overlap'' for indicating the CPF gate characteristics in the main text. Here we show that the entanglement fidelity $F_{e}$ for the CPF gate is consistent with the mode overlap when the input state is equally superposed state.} 

{The entanglement fidelity} $F_{e}$ can be calculated by the maximally entangling states with target qubits $Q$ and ancillary qubits $R$ {(see Ref. [37] in the main text)}.
For the CPF gate in this paper, the coefficients of bases do not mix with each other even if error occurs.
We introduce the wavepacket functions reflected from the atom states $|0\rangle_{\rm{a}}$ ($|1\rangle_{\rm{a}}$) as $f^{0}(x)$ ($f^{1}(x)$), and ideal wavepacket function as $f^{\rm{id}}(x)$ on the basis of the position $x$ on the waveguide, where $\int_{-\infty}^{\infty}dx |f^{\rm{id}}(x)|^{2}=1$.
Note that these functions are different from the $f_{0}(t,\omega)$ and $f_{1}(t,\omega)$ of Eq.~(4) in the main text because, in this case, we do not need to consider states in the cavity and instead consider both polarization states. Hence, the basis is different from that of the main text.
Functions $f^{0}(x)$ and $f^{1}(x)$ are affected by CQED parameters and can be directly calculated by using $L_{0}$ and $L_{1}$ (Eqs.~(5) and (6) in the main text).
Then, an input state $|\Psi_{\rm{in}}\rangle$, an ideal output state $|\Psi_{\rm{id}}\rangle$, a real output state $|\Psi_{\rm{out}}\rangle$, and $F_{e}$ are written as
{\fontsize{10pt}{10pt}\selectfont 
\begin{align}
	|\Psi_{\rm{in}}\rangle
	&=\frac{1}{2}\int_{-\infty}^{\infty}dx 
	f^{\rm{id}}(x)\left(|0H_{x}\rangle_{\rm{Q}}|00\rangle_{\rm{R}}  + |0V_{x}\rangle_{\rm{Q}}|01\rangle_{\rm{R}}+ |1H_{x}\rangle_{\rm{Q}}|10\rangle_{\rm{R}} + |1V_{x}\rangle_{\rm{Q}}|11\rangle_{\rm{R}}\right),\label{S37}\\
	|\Psi_{\rm{id}}\rangle
	&=\frac{1}{2}\int_{-\infty}^{\infty}dx 
	f^{\rm{id}}(x)\left(-|0H_{x}\rangle_{\rm{Q}}|00\rangle_{\rm{R}}  + |0V_{x}\rangle_{\rm{Q}}|01\rangle_{\rm{R}}+ |1H_{x}\rangle_{\rm{Q}}|10\rangle_{\rm{R}} + |1V_{x}\rangle_{\rm{Q}}|11\rangle_{\rm{R}}\right),\label{S38}\\
	|\Psi_{\rm{out}}\rangle
	&=\frac{1}{2}\int_{-\infty}^{\infty}dx 
	\left[-f^{\rm{0}}(x)|0H_{x}\rangle_{\rm{Q}}|00\rangle_{\rm{R}}  + f^{\rm{id}}(x)|0V_{x}\rangle_{\rm{Q}}|01\rangle_{\rm{R}}+ f^{\rm{1}}(x)|1H_{x}\rangle_{\rm{Q}}|10\rangle_{\rm{R}} + f^{\rm{id}}(x)|1V_{x}\rangle_{\rm{Q}}|11\rangle_{\rm{R}}\right],\label{S39}\\
	F_{e}&=|\langle \Psi_{\rm{id}} | \Psi_{\rm{out}} \rangle|^{2} = \left|\frac{1}{4}\left(\int_{-\infty}^{\infty}dx f^{\rm{id}*}(x)f^{\rm{0}}(x)+\int_{-\infty}^{\infty}dxf^{\rm{id}*}(x)f^{\rm{1}}(x)+2\right)\right|^{2},\label{S40}
\end{align}
}
where the basis of qubit states are represented by $|lm_{x}\rangle\equiv|l\rangle_{\rm{a}} \otimes a^{\dagger}_{m}(x)|0\rangle_{\rm{p}}$, (with $l\in\{0,1\}$, $m\in\{H,V\}$), and $a^{\dagger}_{m}(x)$ denoting the creation operator outside of the cavity.
We assume that the vertical polarization remains the ideal wavepacket function since the photon is reflected from the external mirror.

{Next, we also calculate the mode overlap when the input state is the equally surperposition state $|\Psi_{\rm{in}}'\rangle$.}
{Then, an ideal output state $|\Psi_{\rm{id}}'\rangle$ and a real output state $|\Psi_{\rm{out}}'\rangle$ are written as}
{\fontsize{10pt}{10pt}\selectfont 
\begin{align}
	|\Psi_{\rm{in}}'\rangle
	&=\frac{1}{2}\int_{-\infty}^{\infty}dx 
	f^{\rm{id}}(x)\left(|0H_{x}\rangle_{\rm{Q}}  + |0V_{x}\rangle_{\rm{Q}}+ |1H_{x}\rangle_{\rm{Q}} + |1V_{x}\rangle_{\rm{Q}}\right),\\
	|\Psi_{\rm{id}}'\rangle
	&=\frac{1}{2}\int_{-\infty}^{\infty}dx 
	f^{\rm{id}}(x)\left(-|0H_{x}\rangle_{\rm{Q}}  + |0V_{x}\rangle_{\rm{Q}}+ |1H_{x}\rangle_{\rm{Q}} + |1V_{x}\rangle_{\rm{Q}}\right),\\
	|\Psi_{\rm{out}}'\rangle
	&=\frac{1}{2}\int_{-\infty}^{\infty}dx 
	\left[-f^{\rm{0}}(x)|0H_{x}\rangle_{\rm{Q}}  + f^{\rm{id}}(x)|0V_{x}\rangle_{\rm{Q}}+ f^{\rm{1}}(x)|1H_{x}\rangle_{\rm{Q}}+ f^{\rm{id}}(x)|1V_{x}\rangle_{\rm{Q}}\right].
\end{align}
}
The mode overlap between $|\Psi_{\rm{id}}'\rangle$ and $|\Psi_{\rm{out}}'\rangle$ is calculated as
\begin{equation}
	|\langle \Psi_{\rm{id}}' | \Psi_{\rm{out}}' \rangle|^{2} = \left|\frac{1}{4}\left(\int_{-\infty}^{\infty}dx f^{\rm{id}*}(x)f^{\rm{0}}(x)+\int_{-\infty}^{\infty}dxf^{\rm{id}*}(x)f^{\rm{1}}(x)+2\right)\right|^{2}=F_{e}.
\end{equation}
The second equation shows that the mode overlap $|\langle \Psi_{\rm{id}}' | \Psi_{\rm{out}}' \rangle|^{2}$ is consistent with $F_{e}$. 
{In other words, we can calculate $F_{e}$ by inputting only one pure state to the CPF gate.}
This case is unique for the CPF gate, where the coefficients of bases do not mix with each other.
For other general gates, such as CNOT gates or SWAP gates, this does not apply.
Note that this result holds when there is a pure dephasing, as described below.

{
\section{\label{apVI}Effect of pure dephasing}
}
We discuss in the main text the case where a pure-dephasing is negligible. 
This treatment generally holds for CQED systems such as neutral atoms.
In the absence of pure dephasing, a non-Hermitian Hamiltonian method used in this paper and the Lindblad master equation give precisely the same conclusions (see Ref. [11] in the main text). Here, we consider the case where pure dephasing cannot be neglected. In this case, we cannot use a non-Hermitian Hamiltonian method. However, assuming that the pure dephasing is independent of the cavity length, there is no impact on our conclusions about the design of the cavity length, even if the total error rate increases due to the pure dephasing.
In the following, we explain the details.

We can write the ideal output states $|\Psi_{\rm{id}}\rangle$ and the real output states $|\Psi_{\rm{out}}\rangle$ as
{\fontsize{10pt}{10pt}\selectfont 
\begin{align}
	|\Psi_{\rm{id}}\rangle 
	&=\int_{-\infty}^{\infty} dx \left[ -f^{\rm{id}}(x)\alpha_{0H}|0H_{x}\rangle + f^{\rm{id}}(x)\alpha_{0V}|0V_{x}\rangle
	+ f^{\rm{id}}(x)\alpha_{1H}|1H_{x}\rangle + f^{\rm{id}}(x)\alpha_{1V}|1V_{x}\rangle \right] ,
	\label{eq-Psi-id2}\\
	|\Psi_{\rm{out}}\rangle
	&=\int_{-\infty}^{\infty} dx \left[ -f^{\rm{0}}(x)\alpha_{0H}|0H_{x}\rangle + f^{\rm{id}}(x)\alpha_{0V}|0V_{x}\rangle
	+ f^{\rm{1}}(x)\alpha_{1H}|1H_{x}\rangle + f^{\rm{id}}(x)\alpha_{1V}|1V_{x}\rangle \right] ,
	\label{eq-Psi-out2}
\end{align}
}
where $\alpha_{lm}=\alpha_{l}\alpha_{m}$ represents an amplitude of qubit states.
Considering the pure dephasing $e^{\lambda}~(\lambda\leq0)$, the density matrix of the real output state is written as
{\fontsize{7pt}{7pt}\selectfont 
\begin{align}
	&\rho^{\rm{out}} 
	=\int_{-\infty}^{\infty}dx \int_{-\infty}^{\infty}dy \notag\\
	&\left[f^{\rm{0}}(x)f^{\rm{0}*}(y) |\alpha_{\rm{0H}}|^{2} |0H_{x}\rangle\langle 0H_{y}|
	+f^{\rm{0}}(x)f^{\rm{id}*}(y) \alpha_{\rm{0H}}\alpha_{\rm{0V}}^{*} |0H_{x}\rangle\langle 0V_{y}|
	+f^{\rm{0}}(x)f^{\rm{1}*}(y) \alpha_{\rm{0H}}\alpha_{\rm{1H}}^{*} |0H_{x}\rangle\langle 1H_{y}|e^{\lambda}
	+f^{\rm{0}}(x)f^{\rm{id}*}(y) \alpha_{\rm{0H}}\alpha_{\rm{1V}}^{*} |0H_{x}\rangle\langle 1V_{y}|e^{\lambda}\right.\notag\\
	+&f^{\rm{id}}(x)f^{\rm{0}*}(y) \alpha_{\rm{0V}}\alpha_{\rm{0H}}^{*} |0V_{x}\rangle\langle 0H_{y}|
	+f^{\rm{id}}(x)f^{\rm{id}*}(y) |\alpha_{\rm{0V}}|^{2} |0V_{x}\rangle\langle 0V_{y}|
	+f^{\rm{id}}(x)f^{\rm{1}*}(y) \alpha_{\rm{0V}}\alpha_{\rm{1H}}^{*} |0V_{x}\rangle\langle 1H_{y}|e^{\lambda}
	+f^{\rm{id}}(x)f^{\rm{id}*}(y) \alpha_{\rm{0V}}\alpha_{\rm{1V}}^{*} |0V_{x}\rangle\langle 1V_{y}|e^{\lambda}\notag\\
	+&f^{\rm{1}}(x)f^{\rm{0}*}(y) \alpha_{\rm{1H}}\alpha_{\rm{0H}}^{*} |1H_{x}\rangle\langle 0H_{y}|e^{\lambda}
	+f^{\rm{1}}(x)f^{\rm{id}*}(y) \alpha_{\rm{1H}}\alpha_{\rm{0V}}^{*} |1H_{x}\rangle\langle 0V_{y}|e^{\lambda}
	+f^{\rm{1}}(x)f^{\rm{1}*}(y) |\alpha_{\rm{1H}}|^{2} |1H_{x}\rangle\langle 1H_{y}|
	+f^{\rm{1}}(x)f^{\rm{id}*}(y) \alpha_{\rm{1H}}\alpha_{\rm{1V}}^{*} |1H_{x}\rangle\langle 1V_{y}|\notag\\
	+&\left. f^{\rm{id}}(x)f^{\rm{0}*}(y) \alpha_{\rm{1V}}\alpha_{\rm{0H}}^{*} |1V_{x}\rangle\langle 0H_{y}|e^{\lambda}
	+f^{\rm{id}}(x)f^{\rm{id}*}(y) \alpha_{\rm{1V}}\alpha_{\rm{0V}}^{*} |1V_{x}\rangle\langle 0V_{y}|e^{\lambda}
	+f^{\rm{id}}(x)f^{\rm{1}*}(y) \alpha_{\rm{1V}}\alpha_{\rm{1H}}^{*} |1V_{x}\rangle\langle 1H_{y}|
	+f^{\rm{id}}(x)f^{\rm{id}*}(y) |\alpha_{\rm{1V}}|^{2} |1V_{x}\rangle\langle 1V_{y}|\right].
	\label{eq-rho-out}
\end{align}
}
From the discussion in the previous section (VI), the entanglement fidelity $F_{e}$ in the CPF gate can be calculated by inputting the one pure state, i.e., $F_{e} = \langle \Psi_{\rm{id}}| \rho^{\rm{out}}| \Psi_{\rm{id}}\rangle$, where $\rho^{\rm{out}}$ is substituted for $\alpha_{0H}=\alpha_{0V}=\alpha_{1H}=\alpha_{1V}=1/2$. Thus, $F_{e}$ becomes
{\fontsize{8pt}{8pt}\selectfont 
\begin{align}
  F_{e} &=& \langle \Psi_{\rm{id}}| \rho^{\rm{out}}| \Psi_{\rm{id}}\rangle\notag\\
	&=&\frac{1}{4}\int_{-\infty}^{\infty}dx \int_{-\infty}^{\infty}dy \notag\\
	&&\left[ f^{\rm{id}*}(x)f^{\rm{0}}(x)f^{\rm{0}*}(y)f^{\rm{id}}(y)  \right.
	&+&f^{\rm{id}*}(x)f^{\rm{0}}(x) 
	&+&f^{\rm{id}*}(x)f^{\rm{0}}(x)f^{\rm{1}*}(y)f^{\rm{id}}(y)  e^{\lambda}
	&+&f^{\rm{id}*}(x)f^{\rm{0}}(x) e^{\lambda}\notag\\
	&+&f^{\rm{0}*}(y)f^{\rm{id}}(y) 
	&+& 1 
	&+&f^{\rm{1}*}(y)f^{\rm{id}}(y)  e^{\lambda}
	&+& e^{\lambda}\notag\\
	&+&f^{\rm{id}*}(x)f^{\rm{1}}(x)f^{\rm{0}*}(y)f^{\rm{id}}(y)  e^{\lambda}
	&+&f^{\rm{id}*}(x)f^{\rm{1}}(x) e^{\lambda}
	&+&f^{\rm{id}*}(x)f^{\rm{1}}(x)f^{\rm{1}*}(y)f^{\rm{id}}(y) 
	&+&f^{\rm{id}*}(x)f^{\rm{1}}(x) \notag\\
	&+&f^{\rm{0}*}(y)f^{\rm{id}}(y) e^{\lambda}
	&+& e^{\lambda}
	&+&f^{\rm{1}*}(y)f^{\rm{id}}(y)  
	&+&\left. 1 \right].
	\label{eq-rho-out2}
\end{align}
}
Here, the ideal wavepacket function $f^{\rm{id}}(x)$ can be a real number without loss of generality. 
In addition, the discussion in the main text covers the errors by the first-order approximation that includes only photon loss and pulse delay, assuming that the pulse length is long enough for the first-order approximation to be valid. 
Within this approximation, $f^{\rm{0}}(x)$ and $f^{\rm{1}}(x)$  can still be real numbers because they are functions applying photon loss and pulse delay to $f^{\rm{id}}(x)$.
Therefore, from $[f^{\rm{id}}(x), f^{\rm{0}}(x), f^{\rm{1}}(x)]\in \mathbb{R}$, the entanglement fidelity becomes
\begin{align}
  F_{e} &=  \langle \Psi_{\rm{id}}| \rho^{\rm{out}}| \Psi_{\rm{id}}\rangle\notag\\
	&=\frac{1}{4}(z_{0}^{2} + z_{0} 
	+z_{0}z_{1}  e^{\lambda}
	+z_{0}  e^{\lambda}
	+z_{0} 
	+1
	+z_{1} e^{\lambda}
	+  e^{\lambda}\notag\\
	&+z_{0}z_{1} e^{\lambda}
	+z_{1}  e^{\lambda}
	+z_{1}^{2} 
	+z_{1} 
	+ z_{0} e^{\lambda}
	+  e^{\lambda}
	+z_{1}  
	+ 1),\notag\\
	&= \frac{1}{4}[2z_{0}+2z_{1}+z_{0}^{2}+z_{1}^{2}+2+2(z_{0}+z_{0}z_{1}+z_{1}+1)e^{\lambda}].
	\label{eq-rho-out2}
\end{align}
Here, we introduced real numbers $z_{0}\equiv\int_{-\infty}^{\infty}dxf^{\rm{id}}(x)f^{\rm{0}}(x)$ and $z_{1}\equiv\int_{-\infty}^{\infty}dxf^{\rm{id}}(x)f^{\rm{1}}(x)$, where $(z_{0},z_{1})\in \mathbb{R}$ and $0\le(z_{0},z_{1})\le1$.
From $\frac{dF_{e}(z_{0},z_{1})}{dz_{0}}\geq0$ and $\frac{dF_{e}(z_{0},z_{1})}{dz_{1}}\geq0$, 
$z_{0}$ and $z_{1}$ maximizing $F_{e}(z_{0},z_{1})$ are independent of $\lambda$.
Therefore, we can conclude that the cavity parameters that maximize the gate fidelity are independent of pure dephasing under the first-order approximation.


\begin{thebibliography}{10}
\bibitem{note1}
This reason is that one of the conditions for the CPF gate is known as the bandwidth of the photon must be sufficiently smaller than the bandwidth of the cavity~\cite{reiserer2015cavity}.
\bibitem{comm}
Typical CQED systems with neutral atoms have a milisecond order of $T_{2}$~\cite{T2}, and the gate time we analyze is less than microscond order of $\gamma^{-1}$ (see Fig.~\ref{fig-result}). Thus, we neglect the pure dephasing during the CPF gate.
 y
\end{thebibliography}

%




\end{document}